\documentclass[prb,twocolumn,aps,showpacs,fixfloats]{revtex4}
\usepackage{graphicx}
\usepackage{bm}
\usepackage{amsmath,amssymb}
\usepackage{subfigure}
\usepackage{float}
\usepackage{latexsym}
\usepackage{epstopdf}
\usepackage{color}
\usepackage{enumerate}
\usepackage{pdfpages}

\begin{document}
\newcommand{\s}{\scriptscriptstyle}
\newcommand{\uu}{\uparrow \uparrow}
\newcommand{\ud}{\uparrow \downarrow}
\newcommand{\du}{\downarrow \uparrow}
\newcommand{\dd}{\downarrow \downarrow}
\newcommand{\ket}[1] { \left|{#1}\right> }
\newcommand{\bra}[1] { \left<{#1}\right| }
\newcommand{\bracket}[2] {\left< \left. {#1} \right| {#2} \right>}
\newcommand{\vc}[1] {\ensuremath {\bm {#1}}}
\newcommand{\tr}{\text{Tr}}
\newcommand{\Trans}{\ensuremath \Upsilon}
\newcommand{\Refl}{\ensuremath \mathcal{R}}

\title{Spin relaxation of a diffusively moving carrier in a random hyperfine field}

\author{R. C. Roundy  and M. E. Raikh}

\affiliation{ Department of Physics and
Astronomy, University of Utah, Salt Lake City, UT 84112}

\begin{abstract}
Relaxation, $\langle S_z(t)\rangle$, of the average  spin of a carrier in course of hops over sites hosting random hyperfine fields is studied theoretically. In low dimensions, $d=1,2$,  the  decay of average spin with time is non-exponential at all times. The origin of the effect is that for $d=1,2$ a typical random-walk trajectory exhibits numerous self-intersections. Multiple visits of the carrier to the same site accelerates the relaxation since the corresponding partial rotations of spin during these visits add up. Another consequence of self-intersections of the random-walk trajectories is that, in all dimensions, the average,   $\langle S_z(t)\rangle$, becomes sensitive to a weak magnetic field directed along $z$.
Our analytical predictions are complemented by the numerical simulations of  $\langle S_z(t)\rangle$.
\end{abstract}

\maketitle

\noindent {\em Introduction.}
One of the reasons why organic semiconductors are
promising candidates for the active layers of spin valves\cite{v1,v2,v3,v4,v5} is a long spin lifetime, $\tau_s$, in these materials. Due to long $\tau_s$, spin-polarized carriers, injected from one ferromagnetic electrode into the active layer, preserve their spin orientation while traveling towards the other ferromagnetic electrode. As a result, the resistance of the device depends on the mutual orientations of magnetizations of the electrodes (the spin-valve effect). The origin of  slow spin relaxation in organic semiconductors is that they are composed from light atoms with weak spin-orbit coupling.
\begin{figure}[ht]
\includegraphics[width=75mm]{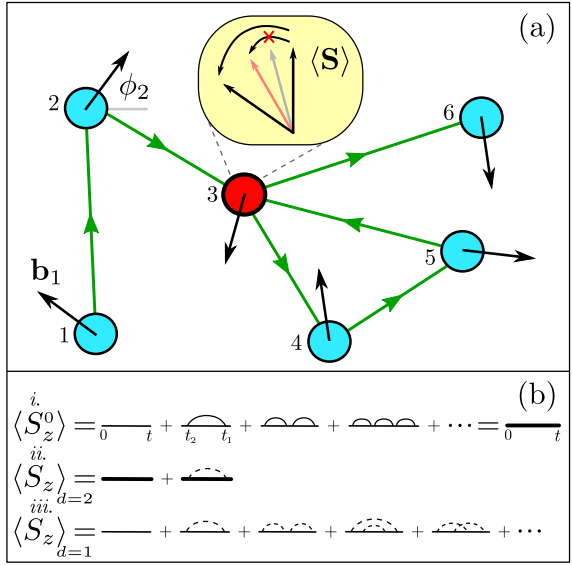}
\caption{(Color online) (a) In course of diffusion
$1\rightarrow 2\rightarrow 3 \rightarrow 4\rightarrow 5\rightarrow 3\rightarrow 6$ over sites hosting random hyperfine fields (black arrows) a carrier visits the site $3$ {\em twice}. As a result, the partial spin rotation doubles, see enlargement; (b) {\em i}. For a trajectory without self-intersections  $\langle S_z(t)\rangle$ is given by
sequence of {\em non-intersecting} solid arcs encoding the correlator $C_0$, {\em ii}. Graphical representation of Eq. (\ref{double-integral}) for the $d=2$ spin  relaxation; self-intersections are captured by a single dashed arc encoding the correlator $C_D$,
{\em iii}. Spin relaxation for $d=1$ is described by diffusive diagrams {\em only}.}
\label{main-figure}
\end{figure}

\begin{figure}[hb]
\includegraphics[width=90mm]{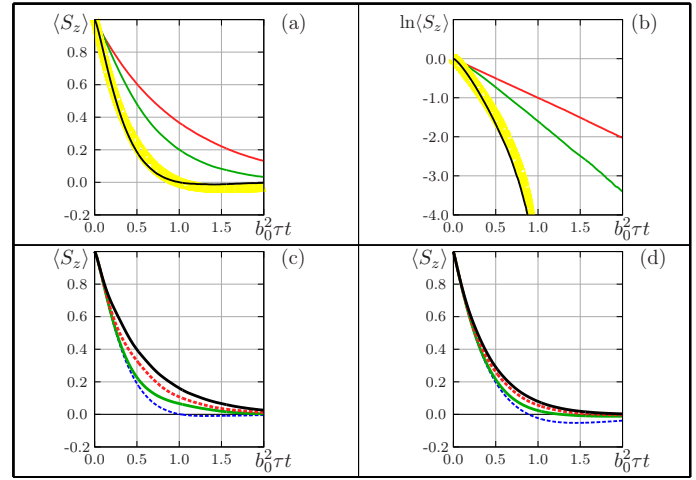}
\caption{ (Color online) (a) $d=2$ spin relaxation for uncorrelated (no self-crossings) hyperfine fields ${\bm b}_i$ (red), with self-intersections and spherically distributed ${\bm b}_i$ (green), and  with self-intersections and planar ${\bm b}_i$ (black); (b) Same as (a) but in log-scale. The decay $\langle S_z(t)\rangle$ is a simple exponent (red), shows crossover between
two simple exponents (green), strongly non-exponential (black).   Yellow line is plotted from Eq. (\ref{double-integral}) with $g_{\s 2} = 0.75$; (c) and (d): weak external field $B\sim \tau_s^{-1}$ suppresses the effect of self-intersections. Numerical (c) and analytical (d) results illustrate how a simple-exponent decay is
restored upon increasing $B\tau_s$. Results for $B=0$ (blue), $B\tau_s=2$ (green), $B\tau_s=5$ (red), and $B\tau_s=10$ (black) are shown. }
\label{2d}
\end{figure}

\begin{figure}[hb]
\includegraphics[width=90mm]{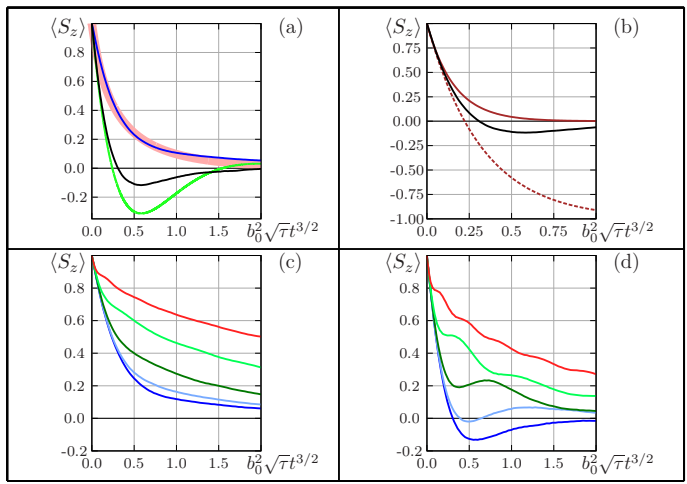}
\caption{(Color online) For $d=1$ random walk the decay, $\langle S_z(t)\rangle$, is a universal function of $b_0^2\sqrt{\tau}t^{3/2}$. (a) numerical results for a planar hyperfine field (black) exhibit spin reversal at intermediate time. For a spherically distributed ${\bm b}_i$ (blue)  the decay is monotonic but non-exponential and is accurately captured by
the solution of the self-consistent equation Eq. (\ref{integro}) (pink). Green curve is the numerical solution of Eq. (\ref{integro}) for planar hyperfine field. (b) Black curve is the same as in (a), while brown and dashed brown show the result of partial summation of the diffusive diagrams, see text.
Weak external field slows down the decay of
$\langle S_z(t)\rangle$ for both spherical (c) and planar  ${\bm b}_i$ (d). Numerical results are shown for the following values of
$\frac{B}{b_0^{4/3}\tau^{1/3}}$: $0$ (dark blue), $1$ (light blue), $2$ (dark green), $4$ (light green), and $8$ (red).}
\label{1d}
\end{figure}

In the absence of spin-orbit coupling, the leading mechanism
of the spin memory loss is a precession of spin in random hyperfine fields created by surrounding  protons on the sites visited by the carrier in course of traveling between the electrodes. With mobility in organic semiconductors being very low, the charge transport in them is via random  inelastic hops of carriers between the sites.
Then the  waiting time, $\tau$, for a subsequent hop plays the role of the correlation time for the random magnetic field, with rms $b_0$, acting on the carrier spin. As a result, the Dyakonov-Perel expression\cite{DP1971} for $\tau_s$ assumes the form
\begin{equation}
\label{tau_s}
\tau_s=\frac{1}{b_0^2\tau}.
\end{equation}
Naturally, for long $\tau_s$, a typical partial  rotation of spin, $\delta\varphi =b_0\tau$, during the waiting time is weak, $\delta\varphi \ll 1$. Assuming that all partial rotations are completely uncorrelated, the spin polarization, averaged over realizations of the hyperfine fields, falls off  with the number
of hops, $N$, as $\langle S_z(N)\rangle
=S_z(0)\exp\left(-N\delta\varphi^2\right)$.
This suggests that the evolution of $\langle S_z\rangle$
with time $t=N\tau$ is a simple exponent
\begin{equation}
\label{exponent}
\langle S_z(t)\rangle=S(0)\exp\Bigl(-\frac{t}{\tau_s}\Bigr).
\end{equation}

The main message of the present paper is that the random walk of a carrier over the sites induces the correlation in hyperfine fields ``sensed" by the  carrier spin. This correlation modifies the decay law Eq. (\ref{exponent}). The origin of correlation  is the self-intersections of the random-walk trajectories, see Fig. \ref{main-figure}. These self-intersections imply multiple visits of the carrier to the {\em same} site. Then the corresponding partial rotations {\em add up} which leads to acceleration of the spin relaxation. The effect is most dramatic if the carrier moves in one dimension. Then, in course of $N$ hops, the carrier  visits  $N^{1/2}$ sites, and the number of visits to a given site is also $N^{1/2}$.  The $N$-dependence of  $\langle S_z\rangle$
can be found from the above derivation of Eq. (\ref{exponent}) upon replacement $N\rightarrow N^{1/2}$ and $\delta\varphi \rightarrow N^{1/2}\delta\varphi$. This yields $\langle S_z(N)\rangle
=S_z(0)\exp\left(-N^{3/2}\delta\varphi^2\right)$, and, correspondingly, the time dependence
\begin{equation}
\label{exponent1}
\langle S_z(t)\rangle=S(0)\exp\Bigl(-\frac{t^{3/2}}{\tau^{1/2}\tau_s}\Bigr).
\end{equation}
In higher dimensions, $d=2$ and $d=3$, the number of self-crossings of an  $N$-step random-walk trajectory is $\sim N$ and $\sim N^{1/2}$, respectively, i.e. each site is visited  twice with probability $\sim 1$ for $d=2$, and with probability $N^{-1/2}$ for $d=3$. As a result, the change, $\langle \delta S_z(t)\rangle$, of the  decay law Eq. (\ref{exponent}) due to accumulation of the partial rotations is of the order of $\langle S_z(t)\rangle$ for $d=2$ and of the order of $\left(\tau/t\right)^{1/2}\langle S_z(t)\rangle$ for $d=3$. But even in the latter case the
correction to Eq. (\ref{exponent}) can be important since it
induces a sensitivity of $\langle S_z(t)\rangle$ to a {\em weak} external magnetic field directed along $z$.
Recall that, without  self-intersections, the $B$-dependence of $\tau_s$ is given by the Hanle-type expression
\begin{equation}
\label{conventional}
\tau_{s}=\frac{1+B^2\tau^2}{ b_{0}^2\tau},
\end{equation}
which applies for $B\gg b_{0}$ and predicts that sensitivity to $B$ emerges at $B\sim \tau^{-1}\gg b_{0}$. We will demonstrate that, with self-crossings of the random-walk trajectories taken into account, the sensitivity to $B$ develops at much smaller field
$B\sim (\tau \tau_s^2)^{-1/3}\ll \tau^{-1}$ in one dimension and at $B\sim \tau_s^{-1}\ll \tau^{-1}$ for $d=2$ and $d=3$. Remarkably, the returns to the same site after
a long time, $t$, give rise to the {\em oscillatory} correction $\propto \cos Bt$ to $\langle S_z(t)\rangle$, which is most pronounced for $d=1$.


\noindent{\em Diagrammatic expansion.}
To illustrate our main message, consider first a simplified situation, when the hyperfine field is located in the $x,y$-plane.  Moreover, we will assume that the randomness in
the in-plane field, $b_\perp = (b_x,b_y)$, is exclusively due
to randomness in the azimuthal angle $\phi$, i.e. $b_x=b_0\cos\phi$, $b_y=b_0\sin\phi$, see Fig.~\ref{main-figure}.

The spin operator satisfies the
equation of motion
$i\frac{d\widehat{S}}{dt} = [\widehat{S}, \widehat{H}]$, with  Hamiltonian
$\widehat{H} =\widehat{S} \cdot {\bm b}(t)$.
Excluding the in-plane components of the operator $\widehat{S}$, the
equation of motion for $S_z$ takes the form
\begin{equation}
\label{equation-of-motion}
S_z(t) = 1 - b_0^2\int\limits_0^t dt_1 \int\limits_0^{t_1} dt_2 \;
\cos (\phi(t_1)-\phi(t_2)) S_z(t_2).
\end{equation}
To find the time evolution of the average, $\langle S_z(t) \rangle$, it is necessary to iterate Eq. (\ref{equation-of-motion}) as
\begin{widetext}
\begin{equation}
\label{expansion}
S_z(t) = 1 - b_0^2\int\limits_0^tdt_1 \int\limits_0^{t_1}dt_2 \cos (\phi(t_1)-\phi(t_2))
+ b_0^4\int\limits_0^tdt_1\int\limits_0^{t_1}\,\,dt_2 \int\limits_0^{t_2}dt_3\int\limits_0^{t_3}dt_4
 \cos (\phi(t_1)-\phi(t_2)) \cos (\phi(t_3)-\phi(t_4))
- \cdots
\end{equation}
\end{widetext}
 and perform averaging over the random azimuthal angle, $\phi(t)$.
Without self-intersections of the random-walk trajectories, this averaging is straightforward since the angles $\phi(t)$,  $\phi(t')$ are correlated only for $|t-t'| \lesssim \tau \ll t$, i.e.
\begin{equation}
\label{tau-correlator}
\langle \cos(\phi(t) - \phi(t')) \rangle = \exp\left[ - |t-t'|/\tau\right]
= C_0(t,t').
\end{equation}
The exponential character of $C_0$ expresses the Poisson distribution of the waiting times.

Each term of the expansion Eq. (\ref{expansion}) can be graphically expressed as a diagram, see Fig. \ref{main-figure}.  Because of the short-time decay of $C_0$, the
arcs corresponding to $C_0$ terms  are not allowed to cross.   More precisely, each crossing of arcs
gives rise to a small factor $\tau/t \ll 1$.
On the other hand, averaging of each term with $n$ non-intersecting arcs yields
$(-1)^n (b_0\tau)^{2n}/n!$, and
we restore Eq. (\ref{exponent}).

As a consequence of self-intersections of the random-walk path, the difference $(\phi(t) -\phi(t'))$ can be small even if the moments $t$ and $t'$ are well separated in time.  Quantitatively, this is captured by the
diffusive contribution, $C_D$, to the correlator

\begin{equation}
\label{diffusive-correlator}
\langle \cos(\phi(t) - \phi(t')) \rangle = \left[ \frac{1}
{2 \pi D |t-t'|} \right]^{d/2}
= C_D(t,t'),
\end{equation}
where the diffusion coefficient $D$ is $1/\tau$ assuming that
the separation between neighboring sites is unity.
In Eq. (\ref{diffusive-correlator}), self-intersections
are accounted for in the continuous limit as a probability
to return to the origin after moving diffusively
for a time $t'-t$.

The correlator $C_D$ should also be incorporated into the
diagrammatic expansion; we denote it with dashed arcs, see
Fig. \ref{main-figure}.  For example, the diagram involving
only one dashed arc is given by
\begin{equation}
\label{one-dashed}
\lambda_d(t)=\hspace{-0.5mm}b_0^2 \int\limits_0^t \hspace{-1mm} dt_1\int\limits_0^{t_1}\hspace{-0.5mm}dt_2
\left[ \frac{1}{2 \pi D \left|t_1-t_2\right|} \right]^{d/2}.
\end{equation}
Evaluation of the double integral yields
\begin{equation}
\lambda_d(t) = \left\{
\begin{matrix}
\dfrac{4}{3(2\pi)^{1/2}} \dfrac{t^{3/2}}{\sqrt{\tau} \tau_s}, & d=1,\\
\dfrac{1}{2 \pi \tau_s}\left[
t \ln\left( \dfrac{t}{\tau} \right) \right],
& d=2,\\
\dfrac{2\tau^{1/2}}{(2\pi)^{3/2} \tau_s} \left(
\dfrac{t}{\sqrt{\tau}} - 2 \sqrt{t}
\right),
& d=3,
\end{matrix}
\right.
\label{lambda-d}
\end{equation}
where we have expressed $b_0^2$ in terms of $\tau_s$ and
have taken into account that the diffusive
description applies when $t_1-t_2 \gtrsim \tau$.
The double integral, Eq. (\ref{one-dashed}) converges
for $d=1$ and the result, Eq. (\ref{lambda-d}),
confirms
the qualitative argument given in the Introduction.
Namely the averaged expansion Eq. (\ref{expansion}) becomes a series in the dimensionless combination $ b_0^2 t^{3/2}\tau^{1/2}$.  Note also, that for $d=1$ the diffusive contribution Eq. (\ref{one-dashed}) exceeds by $(t/\tau)^{1/2}$ the contribution
coming from a single solid-arc.  This illustrates the fact that
each site is visited many times in the course of a $d=1$ random walk.

In two dimensions, the contribution $\lambda_2(t)$, to $\langle S_z(t) \rangle$ from a dashed arc exceeds
logarithmically the contribution from one solid arc.
On the other hand, this contribution contains a prefactor
$(2\pi)^{-1}$. We will take advantage of the smallness
of this prefactor and
sum up {\em all} diagrams containing only
zero or one dashed arc, as illustrated in
Fig. \ref{main-figure}b.   The most delicate ingredient of this
procedure is that 
the insertion of solid arcs under a dashed arc
amounts to the replacement $ C_D(t_1,t_2) \rightarrow C_D(t_1,t_2)
\exp\left[ -\frac{(t_2-t_1)}{2\tau_s} \right]$. Physically, this
means that between the two subsequent visits to the same site at
time moments $t_1$ and $t_2$, the spin polarization is ``forgotten"
in the course of many short-time hops. The emergence of
the  non-trivial factor $1/2$ in the exponent is demonstrated
in the Supplemental Material\cite{Supplemental}where we also show
that the presence of a $z$-component of the hyperfine field amounts
to the replacement $ C_D(t_1,t_2) \rightarrow C_D(t_1,t_2)
\exp\left[ -\frac{3(t_1-t_2)}{4\tau_s}\right]$.

For planar hyperfine fields
the resulting expression for $\langle S_z(t) \rangle$,
which is shown graphically in Fig. \ref{main-figure}b,
takes the form
\begin{equation}
\label{double-integral}
\langle S_z(t) \rangle = e^{-t/\tau_s}
 -\frac{g_{\s 2}}{2 \pi \tau_s}e^{-t/\tau_s}  \int\limits_0^t dt_1
\int\limits_{\tau}^{t_1} dt_2
\frac{\exp\left[ -\frac{(t_2-t_1)}{2\tau_s}\right]}{t_1 - t_2},
\end{equation}
where the numerical factor $g_{\s 2}$ should be $1$,
but is retained intentionally for future comparison with numerics.
The second term is responsible for the deviation from a simple exponential decay.  This term can be easily
reduced to a single integral, and we get
\begin{equation}
\label{Szfor2d}
\langle S_z(t) \rangle = e^{-t/\tau_s}
\left(1- \frac{g_{\s 2}}{\pi}
\int\limits_{\tau/2\tau_s}^{t/2\tau_s}\!\!\frac{dw}{w}
\left( \frac{t}{2\tau_s} - w \right) e^w \right).
\end{equation}
For small $t \ll \tau_s$, Eq. (\ref{Szfor2d})
yields the correction,
$-\frac{g_{\s2}}{2\pi \tau_s} t \ln(t/\tau)$,
to a simple exponent which
reproduces Eq. (\ref{lambda-d}).
In the limit $t \gg \tau_s$ the correction takes
the form
$\frac{g_{\s 2}}{\pi} \left(
\frac{\exp\left(-t/2\tau_s \right)}{(t/2\tau_s)} \right)$.
In fact, this asymptote applies already at $t > \tau_s$.
It decays slower than $\exp(-t/\tau_s)$, so that
$\langle S_z(t) \rangle$ should exhibit a sign reversal
followed by a minimum. Our numerics, see below, shows that
this minimum is very shallow.

Turning now to $d=3$, we find  that the first dominant term in
Eq. (\ref{lambda-d}) describes the contribution from short times
and essentially renormalizes  $\tau_s$. The second subleading term
comes from long diffusive trajectories. It yields a correction to
$\langle S_z(t) \rangle$ which is small as $ \left( \frac{t}{\tau_s} \right)^{1/2}$ at $t\ll \tau_s$ and as $\left(\frac{\tau_s}{t} \right)^{3/2}\exp\left( \frac{-t}{2\tau_s} \right)$ at $t\gg \tau_s$.
The importance of this correction is that it causes a sensitivity of
$\langle S_z(t) \rangle$ to a weak external magnetic field, as we show below.

\noindent {\em Sensitivity to the magnetic field along the $z$-axis.}
Incorporating the constant, ${\bm B}={\bm z}_0B$, and random, $b_z(t)$, components of the magnetic field
amounts to the replacement
\begin{equation}
\Bigl[ \phi(t_1)\!-\!\phi(t_2) \Bigr]\hspace{-1mm} \rightarrow \hspace{-1mm} \left[ \phi(t_1) - \phi(t_2)
+\hspace{-0.3mm} B(t_1\!-\!t_2)\hspace{-0.4mm} + \hspace{-0.5mm}\int\limits_{t_1}^{t_2} dt' b_z(t') \right]
\end{equation}
in Eq. (\ref{expansion}). As discussed in the Introduction, the
solid-arc diagrams describing the hops to nearest neighbors during
the time intervals $\sim \tau$ develop the sensitivity
to $B$ only for strong $B\sim \tau^{-1}$. On the other hand, the dashed-arc
 diagrams are defined by much
longer times, and are thus sensitive to much weaker $B$. The most
interesting domain of $B$ for $d=2,3$ is $\left(b_0\tau\right)^{-1}\gg \frac{B}{b_0}\gg \left(b_0\tau\right)$, where $B\tau$ is small but
$B\tau_s$ is large.  In this domain the decay of $\langle S_z(t)\rangle$ is predominantly exponential while the $B$-dependence comes from the diffusive correction.  For $d=1$ the corresponding domain of $B$ is  $\left(b_0\tau\right)^{-1}\gg \frac{B}{b_0}\gg \left(b_0\tau\right)^{1/3}$. Technically, in this domain,
one can neglect the decoration of the diffusive propagator by solid arcs and rewrite the diffusive contribution Eq. (\ref{one-dashed}) as
\begin{multline}
\label{field-induced}
\delta \langle S_z(t)\rangle
=-\frac{b_0^2\tau^{\frac{d}{2}}}{(2 \pi)^{d/2}}\int\limits_0^tdt_1\int\limits_0^{t_1}dt_2
\frac{\cos B\left(t_1-t_2\right)}{\left(t_1-t_2\right)^{d/2}}\\=
-\frac{b_0^2\tau^{\frac{d}{2}}t^{2-\frac{d}{2}}}{(2\pi)^{d/2}}F_d\left(Bt\right),
\end{multline}
where the function $F_d\left(v\right)$ is defined as
\begin{equation}
\label{Fd}
F_d\left(v\right)=\int\limits_0^1dx\int\limits_0^xdy~\frac{\cos v(x-y)}{\left(x-y\right)^{d/2}}.
\end{equation}
The form of the function, $F_d(v)$, suggests that the diffusive correction contains both  smooth and  {\em oscillating} contributions. The meaning of the smooth contribution
is that,  upon visiting the same site at times $t_1$ and $t_2$, partial spin rotations add up only if $\vert t_1-t_2\vert \lesssim 1/B$. The oscillatory contribution originates from their ``phase shift", $B(t_2-t_1)$.

 Analysis of Eqs. (\ref{field-induced}), (\ref{Fd}) yields the following asymptotes describing the $B$-dependent correction to $\langle S_z(t) \rangle$
\begin{equation}
\label{Fdasymptote}
\left[ F_d(v) - F_d(0)\right]_{v \gg 1} =\varkappa_d(v)
- \frac{\cos v}{v^2},
\end{equation}
where $\varkappa_d(v) = (\pi/2v)^{1/2}$, $-\ln(v)$, and $
-(2 \pi v)^{1/2}$ for $d=1,2,$ and $3$, respectively.
It follows from Eq. (\ref{Fdasymptote}) that magnetic field causes a cutoff of the diffusive correction Eq. (\ref{field-induced}), so that  the value $\delta \langle S_z(t)\rangle$
approaches the value $\sim -b_0^2\tau^{d/2}/B^{2-d/2}$ at large
times. With regard to oscillations, their amplitude falls off with time as $\left(b_0^2/B^2\right)(\tau/t)^{d/2}$, i.e. the oscillations are more pronounced in lower dimensions.

\noindent {\em Numerical results.}
We simulated the spin evolution numerically using the discrete
version of the equation of motion
\begin{multline}
\label{recurrent}
{\bm S}_{i} = \Bigl[ {\bm S}_{i-1} - {\bm n}_i \left(
{\bm n}_i \cdot {\bm S}_{i-1} \right) \Bigr] \cos b_0 \tau \\
+ \left({\bm n}_i \times {\bm S}_{i-1}\right) \sin b_0 \tau
+ {\bm n}_i \left(
{\bm n}_i \cdot {\bm S}_{i-1} \right),
\end{multline}
so that the local hyperfine field had the same magnitude, $b_0$,
on all sites, while the directions, ${\bm n}_i$, were defined by
either a random azimuthal angle, $\phi_i$, or by two spherical angles, $\phi_i$ and $\theta_i$. The diffusive motion of a carrier
was simulated by randomly choosing ${\bm n}_i$ at the next step from
one of the nearest neighbors of ${\bm n}_i$ at the previous step. 

Our numerical results are shown in Figs. \ref{2d} and \ref{1d}.
We started by verifying that, for a directed walk, when {\em all} ${\bm n}_i$ are uncorrelated,  $\langle S_z(i)\rangle$ decays
as a simple exponent.
It is seen from Fig. \ref{2d} that, upon allowing self-intersections, the numerical curve $\langle S_z(t)\rangle$
drops below the result for uncorrelated ${\bm n}_i$ after several
steps. For a spherical hyperfine field, $\ln \langle S_z(t)\rangle$ remains essentially linear at large $t$, but with bigger slope, i.e. the evolution of $\langle S_z(t)\rangle$
exhibits a crossover from one simple exponent at short times to
 another simple exponent at long times. By contrast, for planar hyperfine field,
$\langle S_z(t)\rangle$ is strongly nonlinear in the log-scale at
all times. This completely non-exponential decay
is very well described by Eq. (\ref{Szfor2d}) with
$g_{\s 2} = 0.75$ instead of $1$.

As we argued above, self-intersections give rise to the sensitivity of the spin relaxation to magnetic field $B\sim
\frac{1}{\tau_s}\ll \frac{1}{\tau}$. Evolution of the numerical
curves with $B$ is shown in Fig. \ref{2d}. A significant slowing down  of the relaxation starts from $B\sim \frac{5}{\tau_s}$.
We have also plotted an analytical dependence of $\langle S_z(t)\rangle$ obtained by introducing $\cos Bt(x_1-x_2)$ into
the integrand of Eq. (\ref{double-integral}). Qualitatively, the
numerical and analytical curves exhibit similar behavior. Note however, that the analytical curves saturate at $B\tau_s \sim 15$, while the numerical curves flatten progressively with increasing $B$. This discrepancy is simply due to the fact that the analytical curves correspond to a vanishing product $B\tau$,
and thus cannot capture the conventional spin relaxation
Eq. (\ref{conventional}). With regard to the oscillating correction $\propto \cos Bt$ predicted by  Eq. (\ref{Fdasymptote}), they show up in simulations, but their magnitude is too small to be resolved in Fig. \ref{2d}.

Numerical results for random walk in one dimension are shown in
Fig. \ref{1d}.
Firstly, we established that these numerical results
perfectly satisfy the scaling relation predicted from the qualitative reasoning. Namely, when plotted versus $t^{3/2}b_0^2\tau^{1/2}$, they all fall on a single curve. We also
see that the empirical prediction Eq. (\ref{exponent1}) does not apply. In fact, for
purely planar hyperfine field, the numerical curve, $\langle S_z(t)\rangle$, drops to
a {\em negative} value $\langle S_z\rangle\approx -0.16$ before approaching zero.
As we explained above, only the dashed arcs are responsible for
the spin relaxation for $d=1$. Therefore,
 capturing the nontrivial decay of  $\langle S_z\rangle$ analytically, requires summation of at least a part of dashed-arc-diagrams to {\em all} orders.

In the Supplemental Material we present two variants of such summation.
They essentially reduce to exponentiating of one-dashed-arc
contribution, $\lambda_1(t)$, Eq.~(\ref{one-dashed}),
and differ by the way the numerical factors
in the diagrams with crossings are counted.
Two ways of approximate counting
yield $ \langle S_z(t) \rangle=\exp\left[-\lambda_1(t)\right]$ and
$ \langle S_z(t) \rangle=2\exp\left[-\frac{\lambda_1(t)}{2}\right]-1$, which
lie above and below the numerical results, see Fig. \ref{1d}.
An alternative approach is
to sum only the contributions from non-overlapping diffusive diagrams Fig. \ref{main-figure}.
We show\cite{Supplemental} that this summation leads to a self-consistent equation
\begin{multline}
\label{integro}
\frac{d\langle S_z\rangle}{du}= -\frac{4}{9(2\pi)^{1/2} u^{1/3}}\\
\times \int_0^u \frac{du_1 \exp\left[ - \left( u^{2/3} - u_1^{2/3} \right)^{3/2}  \right ]\langle S_{z} (u_1)\rangle}
{u_1^{1/3} \left(u^{2/3}-u_1^{2/3}\right)^{1/2}},
\end{multline}
where $u=b_0^2\tau^{1/2}t^{3/2}$.
When the hyperfine field is planar, the
numerator in Eq. (\ref{integro}) is $1$.
Eq. (\ref{integro}) does not contain any parameters. Its numerical solutions for spherical and planar hyperfine fields are shown in Fig. \ref{1d}. We see that for spherical case the solution closely reproduces the
simulated decay of  $\langle S_z(t)\rangle$. For planar case, the agreement of the solution with the result of simulation is less accurate. In particular, the depth of the minimum in $\langle S_z(t)\rangle$ predicted by the self-consistent equation is $-0.28$ instead of $-0.16$. Besides, the self-consistent equation predicts additional weak oscillations in $\langle S_z(t)\rangle$ at long times.
Unlike $d=2$, the oscillations in $\langle S_z(t)\rangle$ in
a finite magnetic field are clearly seen in the numerical data, Fig. \ref{1d}. They develop at $B \approx b_0^{4/3}\tau^{1/3}$.

\noindent {\em Discussion.}
Effect of returns to the origin on the spin relaxation was previously discussed in Refs. \onlinecite{KachorovskiiSpinMemory,ShermanNonexponential,SO+1Dlocalization}.
It was assumed that the mechanism of relaxation is
the spin-orbit coupling\cite{DP1971}. For this mechanism, the random field ``sensed'' by electron depends on the direction of its velocity. Then the effect of accumulation of the spin rotation upon multiple visits to the same site, which is central to the present paper, does not apply.
For a unidirectional motion there are no returns and the
{\em average} spin polarization decays as a simple exponent.
 At the same time, the {\em local} spin polarization exhibits very strong fluctuations\cite{we1,we2}.
 To interpret the anomalous sensitivity of $\langle S_z(t)\rangle$ to the external magnetic field, ${\bm B}={\bm z}_0B$, it is instructive to draw analogy to the anomalous sensitivity of the resistance of metals to a  weak
 magnetic field (weak localization\cite{WL}).
 Namely, the phase, $Bt$, of the spin rotation is analogous to the orbital Aharonov-Bohm phase. In weak localization, weak $B$ restricts the area within which counter-propagating random-walk trajectories interfere constructively. In spin relaxation, weak $B$ limits the time interval  within which accumulation of spin rotation due to self-intersections takes place.

\noindent {\em Acknowledgements.}
We are grateful to Sarah Li and Z. V. Vardeny for motivating us.
We are also strongly
grateful to V. V. Mkhitaryan for reading the manuscript
and helpful remarks.
This work was supported by the NSF through MRSEC DMR-112125.

\newpage

\section{Supplemental material}

\subsection{Modification of the diffusive correlator by the short-time
correlators}

In order to establish how the short-time correlations modify
the correlations due to self-intersections of the diffusive
trajectories, consider  the second term of
Eq. (\ref{expansion}).
This term contains the combination
\begin{equation}
\label{FourAngles}
 -\cos \bigl(\phi(t_1)-\phi(t')\bigr) \cos \bigl(\phi(t'')-\phi(t_2)\bigr)
\end{equation}
of four azimuthal angles. Assume that the carrier visits the
same site at distant times moments $t_1$ and  $t_2$, while $t'$ and $t''$ are the initial and final moments of some hop that
takes place between $t'$ and $t''$. Then the averaging over $t'$, $t''$ should be performed using the correlator  Eq.~(\ref{tau-correlator}). To perform this averaging it is
convenient to decompose the product Eq. (\ref{FourAngles})
into a sum

\begin{eqnarray}
-\frac{1}{2} \cos \bigl(\phi(t_1)-\phi(t_2)+ \phi(t'')-\phi(t')\bigr) \nonumber\\
-\frac{1}{2} \cos \bigl(\phi(t_1)+\phi(t_2)- \phi(t')-\phi(t'')\bigr).
\end{eqnarray}
The second term containing the sum, $\bigl(\phi(t')+\phi(t'')\bigr)$,
is zero on average. The first term contains the difference,
$\bigl(\phi(t')-\phi(t'')\bigr)$, and averages to
\begin{equation}
\label{averaged}
 -\frac{1}{2}C_0(t',t'')\cos\bigl(\phi(t_1)-\phi(t_2)\bigr).
\end{equation}
Consider now two intermediate hops taking place within the
intervals $\left[t', t'' \right]$ and, subsequently,  $\left[t''', t'''' \right]$, between the moments $t_1$ and $t_2$.  The corresponding combination to be averaged over the initial
and final moments of the hops reads
\begin{eqnarray}
\cos \bigl(\phi(t_1)-\phi(t')\bigr) \cos \bigl(\phi(t'')-\phi(t''')\bigr)\nonumber\\
\times \cos\bigl(\phi(t'''')-\phi(t_2)\bigr).
\end{eqnarray}
To  average over orientations of the on-site hyperfine fields we have to perform the above decomposition twice, which yields
$\frac{1}{4}C_0(t',t'')C_0(t''',t'''')\cos\bigl(\phi(t_1)-\phi(t_2)\bigr)$.

Upon integration Eq. (\ref{averaged})
over $t'$, $t''$ within the interval $t_1<t'<t''<t_2$, we conclude that a single hop,
described by a solid arc, modifies the integrand in the expression for a diffusive arc by a factor $-\frac{1}{2}\frac{(t_2-t_1)}{\tau_s}$.
Similarly, for two hops, upon integrating over their initial and final moments,
we find that they modify the integrand  by a factor $\frac{1}{4}\frac{(t_2-t_1)^2}{2!\tau_s}$.
Adding the contributions from zero, one, two, three, etc. hops,
we conclude that intermediate hops lead to the factor
$\exp\left[-\frac{(t_2-t_1)}{2\tau_s}\right]$ in the integrand.

In the above reasoning it was implicit that intermediate hops take place
between the time moments $t_1$ and $t_2$. In order to calculate the correction to $\langle S_z(t)\rangle$ due to diffusive random walk, we
also have to take into account the intermediate hops that take place outside the domain $(t_1,t_2)$. In a manner similar to the above, one can
realize that short hops which take place during interval $(0,t_1)$ are accounted for by multiplying $C_D(t_1,t_2)$ by the factor $\exp(-t_1\tau_s)$. We emphasize that, unlike in the domain $(t_1,t_2)$,
the argument of the exponent does not contain $1/2$. Similarly, the short hops in the domain $(t_2,t)$ cause an additional factor $\exp[-(t-t_2)/\tau_s]$. All three factors coming from short hops are taken into account in the second term in Eq. (\ref {double-integral}) of the main text.

In the presence of a $z$-component of the hyperfine field,
the average Eq. (\ref{averaged}) assumes the form
\begin{multline}
\label{averaged1}
 -\frac{1}{2}C_0(t',t'')\cos\bigl(\phi(t_1)-\phi(t_2)\bigr)\\
\times \left< \cos\left( - \int\limits_{t_2}^{t''} b_z(s) ds
- \int\limits_{t'}^{t_1} b_z(s) ds\right) \right>.
\end{multline}
Recall now, that $C_0(t', t'')$ restricts $t'$ and $t''$ within
$\tau$ from each other.  Then the argument of the  cosine in Eq. (\ref{averaged1})
reduces to a single integral, $\Phi = \int\limits_{t_2}^{t_1} b_z(s) ds$. Averaging of this cosine over the realizations of $b_z$
can be performed analytically using the fact that the relevant times $t_1$ and $t_2$ are of the order of $\tau_s$, which is much bigger than $\tau$. Therefore, the phase, $\Phi$, contains many random contributions, which allows us to use the relation
$\langle \cos \Phi \rangle = \exp\left[ - \frac{1}{2} \langle \Phi^2 \rangle \right]$. The average,  $\langle \Phi^2 \rangle$, can be expressed via
$\tau_s$ as $\langle\Phi^2\rangle=\frac{(t_2-t_1)}{4\tau_s}$.
Thus we conclude that, in addition to the factor $-\frac{(t_2-t_1)}{2\tau_s}$,
a single arc brings an additional  factor,
$\exp\left[ - \frac{ (t_1 - t_2)}{4\tau_s} \right]$, into the integrand.

Consideration of two hops
leads to the {\em same} exponential factor,
$\exp\left[ - \frac{ (t_1 - t_2)}{4\tau_s} \right]$.  This follows from the
fact that the product $C_0(t', t'')C_0(t''', t'''')$ restricts $t'$ within $\tau$ from $t''$ and $t'''$ within $\tau$ from $t''''$, which again sets the phase of the cosine,
caused by random $b_z$,
equal to $\int\limits_{t_2}^{t_1} b_z(s) ds$.
Thus, the modification, $\exp\left[ -\frac{3(t_1-t_2)}{4 \tau_s} \right]$, of the integrand of Eq. (\ref{one-dashed})
used in the main text, combines the contributions from in-plane and $z$ components.

\subsection{Partial summation of the diffusive diagrams for $d=1$}

\begin{figure}
\includegraphics[width=77mm]{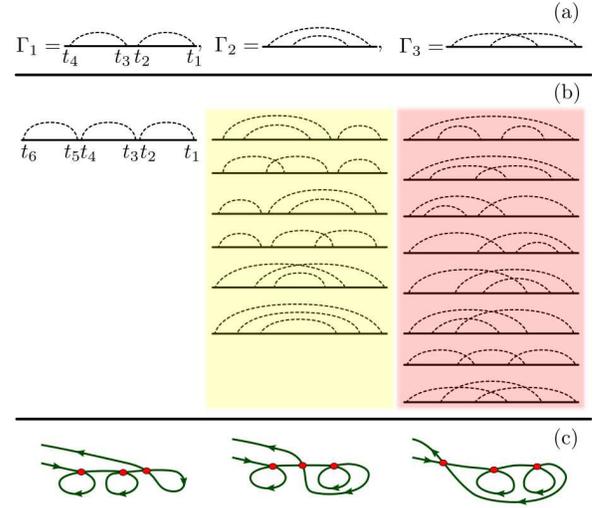}
\caption{(Color online) (a) Three possible diagrams containing two dashed lines. Different pairings
dictate the arguments of the diffusive propagators in the analytical expressions Eq. (\ref{g}) for the
diagrams. (b) Fifteen diagrams with three dashed lines arranged in three groups, white, yellow, and pink, according to their ``complexity".  Different groups correspond to different mutual arrangements
of the self-intersections in the underlying diffusive trajectories. Corresponding trajectories for the
first diagrams of each group are sketched in (c).}
\label{supp-fig1}
\end{figure}

As discussed in the main text, the relevant terms
of expansion Eq. (\ref{expansion}) for $\langle S_z(t)\rangle$ in
one dimension represent only diffusive arcs.
We analyzed only a single-arc contribution, $\lambda_1(t)$, which
is proportional to $b_0^2$. The structure of the two-arc diagrams which are
proportional to $b_0^4$ differs qualitatively from the $b_0^2$-term.
To substantiate this point,
consider a general case of
the term $2n$ in the expansion Eq. (\ref{expansion})
\begin{equation}
\label{Hikami}
\left< \prod\limits_{i,j=1}^{2n}\cos(\phi_i-\phi_j)\right>.
\end{equation}
The average in Eq. (\ref{Hikami})  is nonzero when $\phi_i$ and $\phi_j$ coincide pairwise.
In the simplest case,  $n=1$, the only possible variant of pairing
is $\phi_1=\phi_2$. It corresponds to a single dashed arc.
It is not entirely obvious that, already for $n=2$,
additional variants appear.
Namely, the  product $\cos(\phi_1-\phi_2)\cos(\phi_3-\phi_4)$
survives averaging when $\{\phi_1=\phi_2,~  \phi_3=\phi_4\}$, but also when
$\{\phi_1=\phi_3,~  \phi_2=\phi_4\}$ and $\{\phi_1=\phi_4,~  \phi_2=\phi_3\}$.
As we have shown in the previous subsection, the averaging Eq. (\ref{Hikami})
contains extra $1/2$ for  nontrivial pairings.
Graphically, these nontrivial pairings are described by rainbow-like and crossed
diffusive diagrams, see Fig. \ref{supp-fig1}. Each of three diagrams in Fig. \ref{supp-fig1}
is  $b_0^4\tau t^3$ times a numerical factor.
Denote these  factors with $\Gamma_1$, $\Gamma_2$, and $\Gamma_3$,
so that the arcs in diagram $\Gamma_1$ do not overlap, the arcs in diagram $\Gamma_2$ form a rainbow, while the arcs in diagram $\Gamma_3$ cross.
The explicit expressions for $\Gamma_1$, $\Gamma_2$, and $\Gamma_3$ read
\begin{align}
\label{g}
\Gamma_1 &= 1 \cdot \frac{1}{2\pi}\int\limits_{0}^{1} dx_1 \int\limits_0^{x_1} dx_2
\frac{1}{\sqrt{x_1 - x_2}} \int\limits_0^{x_2} dx_3
\int\limits_0^{x_3} dx_4 \frac{1}{\sqrt{x_3-x_4}}, \nonumber\\
\Gamma_2 &= \frac{1}{2} \cdot \frac{1}{2\pi}\int\limits_{0}^{1} dx_1 \int\limits_0^{x_1} dx_2
\int\limits_0^{x_2} dx_3 \frac{1}{\sqrt{x_2 - x_3}}
\int\limits_0^{x_3} dx_4 \frac{1}{\sqrt{x_1-x_4}}, \nonumber\\
\Gamma_3 &= \frac{1}{2} \cdot \frac{1}{2\pi}\int\limits_{0}^{1} dx_1 \int\limits_0^{x_1} dx_2
 \int\limits_0^{x_2} dx_3 \frac{1}{\sqrt{x_1 - x_3}}
\int\limits_0^{x_3} dx_4 \frac{1}{\sqrt{x_2-x_4}}.
\end{align}
Note now, that if all three  coefficients in front of integrals were
$1/2$, we would be able to present their sum as a single integral
with an integrand being a {\em symmetric} function of all arguments.
This integral can be easily evaluated since it decouples into a
product. Thus we have
\begin{equation}
\label{symmetric}
\Gamma_1+\Gamma_2+\Gamma_3=\frac{\Gamma_1}{2}+\frac{A_1^2}{4}.
\end{equation}
Eq. (\ref{symmetric}) suggests that, upon neglecting $\frac{1}{2}\Gamma_1$,
the sum of the terms containing zero, one, and two dashed arcs can be
presented as $1-\lambda_1(t)+\frac{1}{4}\lambda_1^2(t)$.

We can also look at the sum $\Gamma_1+\Gamma_2+\Gamma_3$ from a
different perspective. Namely, it can be presented as
$\frac{A_1^2}{2}-(\Gamma_2 + \Gamma_3)$,
suggesting that, neglecting $(\Gamma_2 + \Gamma_3)$,
the sum can be presented as $1-\lambda_1(t)+\frac{1}{2}\lambda_1^2(t)$.

The situation with the terms of the order $b_0^6$ offers more options, see Fig. \ref{supp-fig1}.
There is one diagram with a coefficient in front of six-fold integral equal to $1$,
six diagrams with this coefficient equal to  $1/2$, and eight
diagrams with coefficient equal to $1/4$. Again,
if all of the numerical coefficients were the same,
the sum of all $15$ terms reduces to
a single $6$-fold integral with a symmetric integrand,
which can be decoupled into a product of three double
integrals. Namely, if we set all the
 coefficients equal to $1/4$ and neglect
the remainder, the contribution of all $15$
 diagrams would be
equal to $-\frac{1}{4} \left(\frac{A_1^3}{3!}\right)$.
Conversely, if we put all  the coefficients equal to $1$ and neglect the remainder, the result would be $-\frac{A_1^3}{3!}$.

The above arguments can be applied to the higher-order terms proportional to $b_0^{2n}$.
Setting all coefficients equal to $1$, allows to reduce the sum of diagrams with $n$ dashed arcs to $\frac{(-1)^n\lambda_1^n}{n!}$, while setting them all equal to $\frac{1}{2^{n-1}}$ yields $\frac{(-1)^n\lambda_1^n}{2^{n-1}n!}$ for this sum.

Both partial sums can be evaluated analytically.
Namely, the sum is equal to $\exp\left[-\lambda_1(t)\right]$
for the first choice of coefficients
and is equal to $2 \exp\left[-\lambda_1(t)/2 \right] - 1$
for the second one.

The difference between the partial and actual sums
can be estimated by considering the
first neglected terms. Within the first
variant of the partial summation we neglected
$\frac{1}{2}\Gamma_1$, so that
\begin{equation}
\delta \langle S_z \rangle
=  \frac{b_0^4 \tau t^3}{4\pi} \int\limits_0^1\!dx_1
\int\limits_0^{x_1}\!\frac{dx_2}{\sqrt{x_1-x_2}}
\int\limits_0^{x_2}\!dx_3
\int\limits_0^{x_3}\!\frac{dx_4}{\sqrt{x_3-x_4}}.
\end{equation}
The above integral can be evaluated analytically yielding
\begin{equation}
\delta \langle S_z \rangle
=  b_0^4 \tau t^3 \frac{\Gamma_1}{2}
= \frac{1}{24} b_0^4 \tau t^3 .
\label{correction1}
\end{equation}

In a similar fashion we can estimate the accuracy of the
second partial sum in which the first  neglected term is $\Gamma_2+\Gamma_3$. One has
\begin{equation}
\label{correction2}
\delta \langle S_z \rangle
=  \left( \frac{2}{9 \pi} - \frac{1}{24}\right)
b_0^4 \tau t^3
\end{equation}
Note that, while the two corrections Eq. (\ref{correction1}) and
Eq.~(\ref{correction2}) are almost equal in magnitude, the first
one shifts the partial sum, $2\exp{\left[-\lambda_1(t)/2\right]}-1$, up, while the second one shifts the partial sum, $\exp{\left[-\lambda_1(t)\right]}$, down.

\subsection{Self-consistent equation for $\langle S_z(t)\rangle$ }

Summation of the subset of diagrams with $n$ non-overlapping dashed arcs, Fig. \ref{supp-fig1}, corresponds to retaining only the pairings $\phi(t_1)=\phi(t_2)$,~ $\phi(t_3)=\phi(t_4)$, .... in Eq. (\ref{expansion}).
With these pairings, upon  averaging of  Eq. (\ref{expansion}) over
the hyperfine fields with the help of Eq. (\ref{diffusive-correlator}),
the expression for the average $\langle S_z(t)\rangle$
assumes the form
\begin{equation}
\label{1}
\langle S_z(t)\rangle=\sum_{n=0}^\infty(-1)^n b_0^{2n}\mu^{(n)}(t),
\end{equation}
where $\mu^{(n)}$ are the $2n$-fold integrals defined as
\begin{equation}
\label{2}
\mu^{(n)}(t)=\int\limits_0^t dt_1 \int\limits_0^{t_1} dt_2 C_D(t_1,t_2)
\int\limits_0^{t_2} dt_3 \int\limits_0^{t_3}dt_4 C_D(t_3,t_4)....
\end{equation}
It is apparent from Eq. (\ref{2}) that these integrals satisfy the recurrence relation
\begin{equation}
\label{3}
\frac{d\mu^{(n+1)}}{dt}=
\int\limits_0^t dt_1 C_D(t,t_1)\hspace{0.5mm}\mu^{(n)}(t_1).
\end{equation}
Using this relation, the derivative $d\langle S_z\rangle/dt$ can be expressed
via $\langle S_z\rangle$ as follows
\begin{equation}
\label{4}
\frac{d\langle S_z\rangle}{dt}
=-
\frac{b_0^2\tau^{1/2}}{(2\pi)^{1/2}}
\int\limits_0^t \frac{dt'}{\left(t-t'\right)^{1/2}}\langle S_z(t') \rangle,
\end{equation}
where we have substituted the explicit form of the $1d$ diffusive correlator.
Upon introducing a new variable
$u=b_0^2\tau^{1/2}t^{3/2}$,
Eq. (\ref{4}) can be reduced to a dimensionless form
\begin{equation}
\label{5}
\frac{d\langle S_z\rangle}{du}=-\frac{4}{9(2\pi)^{1/2}u^{1/3}}\int_0^u\frac{du_1 }{u_1^{1/3}\left(u^{2/3}-u_1^{2/3}\right)^{1/2}}\langle S_z(u_1)\rangle.
\end{equation}
Numerical solution  of Eq. (\ref{5}) exhibits rapidly decaying oscillations at $u\gg 1$.
Only the first oscillation can be resolved in Fig. \ref{1d}a. The depth of a corresponding minimum
is $\langle S_z\rangle =-0.34$, i.e. it is approximately two times deeper than the minimum in the 
simulation result.    


In the presence of $b_z$-component of the random hyperfine field  $C_D(t_1,t_2)$ in Eq. (\ref{2})
gets modified as
\begin{equation}
C_D(t_1, t_2) \rightarrow C_D(t_1, t_2) \left< \cos \left( - \int\limits_{t_2}^{t_1} b_z(s) ds\right) \right>.
\end{equation}
Averaging of the cosine again reduces to the exponent
\begin{multline}
\label{b_z}
\left< \cos \left( - \int\limits_{t_2}^{t_1} b_z(s) ds\right) \right>=\exp\Bigl[-\frac{1}{2}\int\limits_{t_1}^{t_2}
\int\limits_{t_1}^{t_2}dt'dt''\langle b_z(t')b_z(t'')\rangle\Bigr]\\=\exp\Bigl[-\frac{b_0^2}{4}\int\limits_{t_1}^{t_2}
\int\limits_{t_1}^{t_2}dt'dt''C_D(t',t'')\Bigr].
\end{multline}
Note that, unlike modification of a single dashed arc, the exponent contains
diffusive correlator, $C_D(t',t'')$, rather than the short-time
correlator $C_0(t',t'')$.
With the above modification of $C_D(t_1,t_2)$,  the self-consistent equation Eq.~(\ref{4}) takes the form
\begin{equation}
\label{8}
\frac{d\langle S_z\rangle}{dt}
=-
\frac{b_0^2\tau^{1/2}}{(2\pi)^{1/2}}
\int\limits_0^t \frac{dt'\exp\left[-b_0^2\tau^{1/2}(t-t')^{3/2}\right]}{
\left(t-t'\right)^{1/2}}\langle S_z(t') \rangle,
\end{equation}
which leads to Eq. (\ref{integro}) of the main text.
\end{document}